\documentclass[12pt,pagenumbers]{spieman}  % 12pt font required by SPIE;
\usepackage{amsmath,amsfonts,amssymb}
\usepackage{graphicx}
\usepackage{setspace}
\usepackage{tocloft}
\usepackage{lineno}
\usepackage[misc]{ifsym}

\title{Distribution-informed and wavelength-flexible data-driven photoacoustic oximetry}

\author[1, 2, \Letter]{Janek Gr\"ohl}
\author[1, 2]{Kylie Yeung} % kylie.yeung@univ.ox.ac.uk
\author[1, 2]{Kevin Gu} % not yet in contact
\author[1, 2]{Thomas R. Else}
\author[1, 2]{Monika Golinska}
\author[1, 2]{Ellie V. Bunce}
\author[3]{Lina Hacker}
\author[1, 2, \Letter]{Sarah E. Bohndiek}

\affil[1]{Cancer Research UK Cambridge Institute, University of Cambridge, Cambridge, UK}
\affil[2]{Department of Physics, University of Cambridge, Cambridge, UK}
\affil[3]{Department of Oncology, University of Oxford, Oxford, U.K.}
\cftpagenumbersoff{figure}
\cftpagenumbersoff{table}

\begin{document} 
\maketitle
%\linenumbers

\begin{abstract}

\textbf{Significance}: Photoacoustic imaging (PAI) promises to measure spatially-resolved blood oxygen saturation, but suffers from a lack of accurate and robust spectral unmixing methods to deliver on this promise. Accurate blood oxygenation estimation could have important clinical applications, from cancer detection to quantifying inflammation.\\

\textbf{Aim}: This study addresses the inflexibility of existing data-driven methods for estimating blood oxygenation in PAI by introducing a recurrent neural network architecture. \\

\textbf{Approach}: We created 25 simulated training dataset variations to assess neural network performance. We used a long short-term memory network to implement a wavelength-flexible network architecture and proposed the Jensen-Shannon divergence to predict the most suitable training dataset.\\

\textbf{Results}: The network architecture can handle arbitrary input wavelengths and outperforms linear unmixing and the previously proposed learned spectral decolouring method. Small changes in the training data significantly affect the accuracy of our method, but we find that the Jensen-Shannon divergence correlates with the estimation error and is thus suitable for predicting the most appropriate training datasets for any given application.\\

\textbf{Conclusions}: A flexible data-driven network architecture combined with the Jensen-Shannon Divergence to predict the best training data set provides a promising direction that might enable robust data-driven photoacoustic oximetry for clinical use cases.

\end{abstract}

% Include a list of up to six keywords after the abstract

{\noindent \footnotesize{\Letter} Send correspondence to \linkable{jmg236@cam.ac.uk, seb53@cam.ac.uk} }

\begin{spacing}{1}   % use double spacing for rest of manuscript

\section{Introduction}
\label{sec:intro}  % \label{} allows reference to this section

% The unique selling point of PAI is spatially-resolved oximetry
Blood oxygen saturation (sO$_2$) is an important indicator of individual health status used routinely in patient management~\cite{fawzy2022racial}. Photoacoustic imaging (PAI) is a promising medical imaging modality for real-time non-invasive  spatially-resolved measurement of sO$_2$~\cite{beard2011biomedical}, with early clinical applications~\cite{li2018photoacoustic} shown in, for example, inflammatory bowel disease~\cite{knieling2017multispectral} and cardiovascular diseases~\cite{karlas2019cardiovascular}. In cancer, alterations in localized sO$_2$ levels have been linked to angiogenesis and hypoxia~\cite{dewhirst2008cycling}; key hallmarks of cancer that are known to affect treatment outcomes~\cite{hanahan2022hallmarks} and are measurable through PAI. \\

% Linear unmixing is typically used but is not ideal
Unfortunately, it remains difficult to apply PAI to derive quantitative values for sO$_2$ from multispectral PA measurements~\cite{laufer2006quantitative, bench2021quantitative}. Linear unmixing (LU) remains the \textit{de facto} standard for sO$_2$ estimation from PAI measurements~\cite{tzoumas2017spectral} because of its simplicity and flexibility, but has well-understood limitations in its applicability and accuracy~\cite{hochuli2019estimating}. The limitations of LU are significant in the context of artefacts arising from: the optical processes (e.g. non-linear light fluence distribution~\cite{cox2012quantitative} leading to spectral colouring), acoustic processes (e.g. reflection artefacts~\cite{singh2016vivo}), or reconstruction algorithms (e.g. model mismatch in sound speed~\cite{jose2012speed}). Using LU is particularly challenging in the presence of highly absorbing tissues, such as the epidermis,~\cite{mantri2022impact} which can introduce reflection artefacts and a spectral bias that leads to an overestimation of sO$_2$, which increases with darker skin tone~\cite{else2023effects}.\\

% Data-driven approaches show promise but are difficult to validate and can be unpredictable when applied to new datasets
Data-driven unmixing schemes have shown promise to alleviate some of the shortcomings of LU~\cite{luke2019net,agrawal2021learning,grasso2022superpixel} but suffer from three major drawbacks: (1) inflexibility to receiving different input data after training~\cite{grohl2021learned}; (2) performance determined by the composition of the training dataset~\cite{grohl2021deep}; and (3) limited testing on diverse and representative use cases~\cite{assi2023review}. In comparison to LU, data-driven methods are often inflexible regarding the input data after training, impacting generalizability, making them difficult to use, and requiring laborious tailoring to a specific application and imaging system. Thus, data-driven sO$_2$ estimation methods can struggle to translate from promising findings \textit{in silico} to \textit{in vivo} data~\cite{sandfort2019data,maier2022surgical}. The lack of high-quality annotated training data and reliable validation data has made it difficult to implement data-driven methods robustly. One way of tackling this challenge lies in bridging the gap between simulated and actual PA data, exploiting realistic phantoms~\cite{hacker2022criteria}, an approach that has recently started to be explored~\cite{grohl2023moving,susmelj2023signal,dreher2023unsupervised}. Further, many data-driven approaches use single-pixel input spectra for their inversion, as with LU, even though 3D inversions would be preferred for realistic use cases~\cite{bench2020toward}. Solving the full 3D problem is typically computationally intensive, limiting its success \textit{in vivo}~\cite{cox2009estimating}. To make the inverse problem more tractable without full 3D information, some approaches use priors in the inversion scheme~\cite{tzoumas2016eigenspectra}, differential image analysis~\cite{bench2021quantitative}, or multiple measurements with differences in illumination~\cite{kirchner2021multiple}.\\

% MAIN CONTRIBUTIONS:
In this work, we set out to address the three aforementioned limitations. We improve the flexibility of data-driven sO$_2$ estimation using a Long Short Term Memory (LSTM) network that enables input wavelength flexibility. We propose a method to inform the choice of training data, which can either be used to choose the best pre-trained model for a target application or to inform the choice of simulation parameters when creating a training data set to underpin a new model. We test these methods on diverse data sources across simulations, phantoms, small animals and humans.

\section{Materials and Methods}
\label{sec:methods}

\begin{figure}[h!tb]
    \centering
    \includegraphics[width=\textwidth]{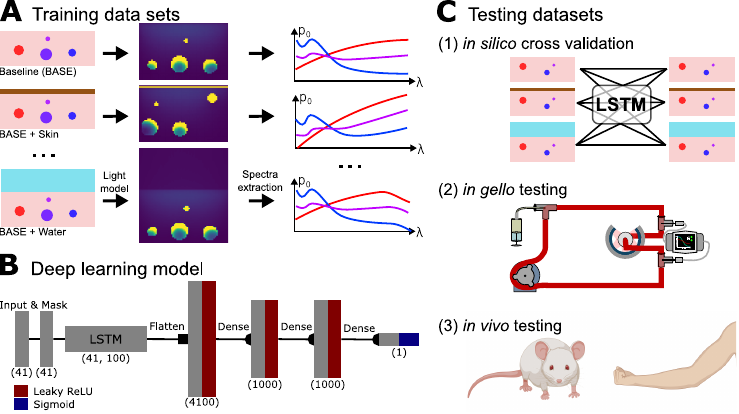}
    \caption{\textbf{Overview of the methods used.} \textbf{A} Photon transport was simulated with a Monte Carlo light model for each of 25 distinct datasets adapted from a baseline tissue assumption (BASE). Spectra with simulated amplitude for each wavelength were extracted from these simulations. \textbf{B} A deep learning network based on a long short-term memory (LSTM) network was introduced to enable greater flexibility regarding input wavelengths for analysis. The hidden state of the LSTM was passed to fully connected layers, which output the estimated blood oxygenation sO$_2$. \textbf{C} The performance of the LSTM-based method when trained on datasets with different tissue simulation parameters was tested across different datasets, ranging from \textit{in silico} simulations, \textit{in gello} phantom measurements, to \textit{in vivo} measurements. This figure was created with Inkscape using BioRender assets.}
    \label{fig:method-overview}
\end{figure}

We begin by investigating sensitivity to changes in training dataset simulation parameters, by defining a baseline dataset set (BASE) with typical assumptions on the tissue geometry and functional parameter ranges, then adapt it into 25 variations (Fig \ref{fig:method-overview} A). We use the Jensen-Shannon divergence to determine the ideal training dataset for a given use case. We propose a deep learning network architecture based on a long short-term memory (LSTM) network that is flexible regarding the input wavelengths (Fig \ref{fig:method-overview} B) and use a testing strategy that comprises computational studies \textit{in silico}, phantoms \textit{in gello}~\cite{manohar2023gello}, and \textit{in vivo} data (Fig \ref{fig:method-overview} C).

\subsection{\textit{In silico} datasets}

25 \textit{in silico} datasets were simulated in Python using the SIMPA toolkit~\cite{grohl2022simpa} (Fig\ref{fig:method-overview} A). A Monte Carlo model (MCX~\cite{fang2009monte}) was used for the optical forward model with a 50\,mJ Gaussian beam using $10^7$ photons and a 20\,mm radius, simulating at 41 wavelengths from 700\,nm to 900\,nm in 5\,nm steps, and assuming an anisotropy of 0.9. For the datasets that include acoustic modelling, a 2D k-space pseudo-spectral time domain method implemented in the k-Wave toolbox~\cite{treeby2010k} was used. We assumed a uniform speed of sound of 1500\,ms$^{-1}$, a density of 1000\,gcm$^{-3}$, and disregarded acoustic absorption. For most datasets, a generic linear ultrasound detector array was placed in the centre of the simulated volume. The generic array consists of 100 detection elements with a pitch of 0.18\,mm, a centre frequency of 4\,MHz, a bandwidth of 55\%, and a sampling rate of 40\,MHz. For the datasets mimicking the setup of commercial instruments, we used the built-in device definitions provided in SIMPA.\\

\textbf{Baseline dataset parameters.} The parameters of the baseline dataset (BASE) were chosen to reflect the typical parameter choices found in literature~\cite{grohl2021learned, cai2018end, chen2020deep, bench2020toward, hoffer2019absO2luteu, olefir2020deep}. A 19.2 x 19.2 x 19.2\,mm cube was simulated with a 0.3\,mm voxel size, resulting in $64^3$ voxels per volume. The background started from the second voxel from the top and was modelled as muscle tissue with a blood volume fraction of 1\%~\cite{grohl2021learned} with 70\% oxygenation~\cite{lyons2016mapping}. We added 0-9 cylinders with a diameter randomly chosen from U[0.3\,mm, 2.0\,mm], where U denotes a uniform distribution. The tissue was divided into 3 x 3 equal-sized compartments in the imaging plane and a vessel was included with a 50\% probability. Each vessel purely contained blood with a randomly drawn sO$_2$ value from U[0\%, 100\%]. To prevent vessel overlap, they were located within their respective compartments' boundaries.\\

We defined 24 variations of BASE (Table~\ref{tab:datasets}). The variations encompassed a range of tissue backgrounds, vessel sizes, illumination geometries and resolutions, as well as the addition of a skin layer, performing acoustic simulation, and using digital twins of two commercial instruments (MSOT Acuity Echo and  MSOT InVision 256-TF; iThera Medical GmbH, Munich, Germany). The MSOT Acuity Echo is a handheld clinical PAI device with 256 detection elements and an angular coverage of 120$^\circ$, whereas the MSOT InVision 256-TF is a tomographic pre-clinical PAI device with 256 detection elements and a 270$^\circ$ view angle. The full details for the digital twin parameters of these devices can be found in prior work~\cite{grohl2022simpa,grohl2023moving} and are available within the SIMPA toolkit.\\

\begin{table}[h!tb]
    \centering
    \begin{tabular}{rp{11cm}}
    \textbf{Dataset identifier} & \textbf{Changes from BASE}\\
    \hline
    \\
    BG\_0-100 & Background sO$_2$ randomly chosen from U(0\%, 100\%). \\
    BG\_60-80 & Background sO$_2$ randomly chosen from U(60\%, 80\%).\\
    BG\_H2O & Background modelled as water only.\\
    HET\_0-100 & Background heterogeneously varied between 0\% and 100\%.\\
    HET\_60-80 & Background heterogeneously varied between 60\% and 80\%.\\
    RES\_0.15 & Simulation grid spacing: 0.15\,mm.\\
    RES\_0.15\_SMALL & Simulation grid spacing: 0.15\,mm; radii of vessels halved.\\
    RES\_0.6 & Simulation grid spacing: 0.6\,mm.\\
    RES\_1.2 & Simulation grid spacing: 1.2\,mm.\\
    SKIN & 0.3\,mm melanin layer; melanosome fraction: U(0.1\%, 5\%). \\
    ILLUM\_5mm & Radius of incident beam: 5\,mm.\\
    ILLUM\_POINT & Radius of incident beam: 0\,mm.\\
    SMALL & Radii of the vessels halved.\\
    ACOUS & Acoustic modelling with linear transducer array.\\
    WATER\_2cm & 2cm H$_2$O layer added between illumination and tissue.\\
    WATER\_4cm & 4cm H$_2$O layer added between illumination and tissue.\\
    MSOT & Volume extended: 75 x 19.2 x 19.2\,mm; MSOT Acuity twin.\\
    MSOT\_SKIN & MSOT + SKIN. \\
    MSOT\_ACOUS & MSOT + ACOUS.\\
    MSOT\_ACOUS\_SKIN & MSOT + SKIN + ACOUS.\\
    INVIS & Volume extended: 90 x 25 x 90\,mm; MSOT InVision 256-tf twin. Single vessel in a 10\,mm radius tubular background.\\
    INVIS\_SKIN & INVIS + SKIN\\
    INVIS\_ACOUS & INVIS + ACOUS.\\
    INVIS\_SKIN\_ACOUS & INVIS + SKIN + ACOUS.\\
    \\
    \end{tabular}
    \caption{Summary of datasets used in the study. Each row identifies the changes in dataset creation parameters performed for each of the training datasets relative to the BASE dataset (specified in Methods subsection 2.1). The left column shows an identifier assigned to each dataset that is used throughout the manuscript and the right column summarises the deviation from BASE. U denotes a uniform distribution.}
    \label{tab:datasets}
\end{table}

\textbf{Data preprocessing.} We extracted pixel-wise PA spectra from simulations of initial pressure or reconstructed signal amplitude if acoustic simulations were performed. We only extracted spectra from blood vessels where the signal intensity at 800\,nm was at least 10\% of the maximum signal intensity in the dataset. If less than 10\% of voxels were chosen this way, we selected the 10\% of voxels with the highest signal amplitude from the dataset. We enforced this selection criterion to effectively exclude voxels with low signal-to-noise ratio (SNR) caused by optical attenuation, as previous works have shown that idealised simulation can contain spectra where the spectral colouring is unrealistically strong in regions with low signal amplitude~\cite{grohl2021learned, kirchner2021multiple}.\\

The number of extracted spectra from the different dataset variations ranged from 25 thousand to 60 million, compared to BASE with 7 million spectra; the mean over all datasets was just above 6 million. The large difference in extractable spectra is primarily caused by two factors: (1) typical simulations yield 3D p$_0$ distributions, but adding acoustic forward modelling leads to a 2D reconstructed image; and (2) some datasets only include a single vessel in the centre of the phantom tube, mimicking a blood flow phantom~\cite{gehrung2019development} (described below). To mitigate performance differences caused by this discrepancy, we stratified the dataset sizes by randomly sampling 300,000 spectra with replacement per dataset, which represents a balanced compromise between undersampling larger datasets and oversampling smaller ones~\cite{halevy2009unreasonable}.\\

We performed z-score normalization on each spectrum, setting the mean ($\mu$) to 0 and variance ($\sigma^2$) to 1, which discards signal intensity information and eliminates the need for quantitative simulation calibration. Since we performed the same spectra-wise z-score normalisation on experimental data, this normalisation allows us to apply sO$_2$ estimation algorithms trained \textit{in silico} to experimental \textit{in gello} and \textit{in vivo} data.\\

\subsection{Deep Learning Algorithm}

To address the limited flexibility of data-driven oximetry methods~\cite{olefir2020deep,grohl2021learned}, a custom unmixing network architecture was composed that contains a Long Short-Term Memory (LSTM) network. Due to the recurrent nature of an LSTM, it can process sparse spectra containing zeros at arbitrary positions (see Fig~\ref{fig:method-overview} B).\\

The network input size was fixed at 41, representing the maximum number of wavelengths (between 700\,nm and 900\,nm in 5\,nm steps) we consider during inference. The number is a trade-off between maximising the spectral resolution of the input features and simulation time efficiency. The network started with an LSTM layer with a hidden size of 100. A masking layer was introduced to identify zeroed values, creating a mask that instructs the layer to ignore missing values. The LSTM output was then flattened into a fixed-length encoding. Following the LSTM, a three-layer fully connected neural network (FCNN) was used with input size 4200, hidden size 1000, and output size 1. A leaky rectified linear unit was used after each layer and the activation function after the final layer was a sigmoid function to constrain sO$_2$ predictions between 0 and 1 (Fig~\ref{fig:method-overview} B).\\

The deep learning networks were trained for 100 epochs, where each epoch included the entire training set. The parameters were optimised with the Adam optimizer from the Keras framework~\cite{chollet2015keras} using an initial learning rate of 10$^{-3}$ and a mean absolute error loss. The learning rate was halved upon a 5-epoch plateau of the validation loss.

\subsection{\textit{In gello} blood flow phantom imaging}

Two variable oxygenation blood flow phantoms were imaged using a previously described protocol~\cite{gehrung2019development}. Briefly, agar-based cylindrical phantoms with a radius of 10\,mm were created and a polyvinyl chloride tube (inner diameter 0.8\,mm, outer diameter 0.9\,mm) was embedded in the centre before the agar was allowed to set at room temperature. The base mixture comprised 1.5 \% (w/v) agar (Sigma Aldrich 9012-36-6) in Milli-Q water and was heated until dissolved. After cooling to approximately 40$^\circ$\,C, 2.08\% (v/v) of pre-warmed intralipid (20\% emulsion, Merck, 68890-65-3) and 0.74\% (v/v) Nigrosin solution (0.5 mg/mL in Milli-Q water) were added, mixed, and poured into the mold. Imaging was performed using the MSOT InVision 256TF (iThera Medical GmbH, Munich, Germany) according to a previously established standard operating procedure \cite{joseph2017evaluation}. PA images of the phantom were acquired in the range between 700\,nm and 900\,nm with 20\,nm increments. The first phantom was imaged using deuterium oxide (D$_2$O, heavy water) as the coupling medium within the system, while the second was immersed in normal water (H$_2$O) during imaging. 

\subsection{In vivo human forearm imaging}

Human forearm imaging was performed as part of the PAISKINTONE study, which started in June 2023 following approval by the East of England - Cambridge South Research Committee (Ref: 23/EE/0019). The study was conducted in accordance with the Declaration of Helsinki and written informed consent was obtained from all study participants. Participants were excluded if they could not give consent, were under the age of 20 or over 80, or had a body mass index (BMI) outside the range between 18.5 and 30. Imaging was performed using the MSOT Acuity Echo (iThera Medical GmbH, Munich, Germany) using laser light between 660\,nm and 1300\,nm, averaging over 10 scans each and analysis was performed at five wavelengths (700\,nm, 730\,nm, 760\,nm, 800\,nm, 850\,nm). One forearm scan from N=7 randomly chosen subjects with Fitzpatrick Type 1 or 2 were selected for the purposes of testing the method proposed in this work. The authors manually segmented the radial artery in each scan using the medical imaging interaction toolkit (MITK)~\cite{wolf2005medical}.

\subsection{\textit{In vivo} mouse imaging}

All animal procedures were conducted under project and personal licences (PPL no PE12C2B96, PIL no I53057080), issued under the United Kingdom Animals (Scientific Procedures) Act, 1986 and compliance was approved locally by the CRUK Cambridge Institute Biological Resources Unit. Nine healthy 9-week-old female C57BL/6 albino mice were imaged using the MSOT InVision 256TF (iThera Medical GmbH, Munich, Germany) according to a previously established standard operating procedure \cite{joseph2017evaluation}. Additionally, six 28-week-old healthy female BALB/c nude mice were imaged while inhaling 100\% CO$_2$ as their terminal procedure. In both cases, imaging was performed at 10 wavelengths equally spaced between 700 and 900\,nm averaging over 10 scans each. The mouse body, kidneys, spleen, spine, and aorta were manually segmented by the authors using MITK.

\subsection{Performance evaluation}

The performance of the LSTM-based method was evaluated using the median absolute error ($\epsilon \text{sO}_2$) between the estimate ($\hat{\text{sO}_2}$) and the ground truth/reference sO$_2$:

\begin{equation}
    \epsilon \text{sO}_2 = \text{median}(|\text{sO}_2 - \hat{\text{sO}_2}|)
\end{equation}

Ground truth values are available for the \textit{in silico} and \textit{in gello} datasets. For the \textit{in vivo} measurements, reference values were based on literature. We assumed sO$_2$ of mixed murine blood to be 60\% - 70\%~\cite{lyons2016mapping} and of arterial murine blood sO$_2$ under anaesthesia to be 94\% - 98\%~\cite{loeven2018arterial}. For the CO$_2$ terminal procedure, we assumed that CO$_2$ binds to haemoglobin, forming carbaminohaemoglobin, which leads to oxygen unloading~\cite{abramczyk2023hemoglobin} and has an absorption spectrum similar to deoxyhaemoglobin~\cite{dervieux2020measuring,taylor2022noninvasive}, thus continuously decreasing the actual~\cite{huber2002co2} and measured global blood sO$_2$. In humans, arterial blood sO$_2$ of 95\%-100\% was assumed~\cite{williams1998assessing}.

\subsection{Simulation gap measure}
\label{sec:djs}

To predict the best-fitting training data set for a target application, one could use the sO$_2$ estimates of a trained algorithm to calculate error metrics, such as the absolute estimation error, but this is only possible when ground truth or reference sO$_2$ values for a representative dataset are available. For \textit{in vivo} applications this is typically not the case and additionally, unsupervised methods for performance prediction in the context of PA oximetry remain largely unexplored.

Our alternative solution to this problem uses the Jensen-Shannon divergence~\cite{lin1991divergence} (D$_\text{JS}$). To this end, we compute D$_\text{JS}$ between the training data distributions and the target data distribution and calculate the Pearson correlation coefficient between the resulting D$_\text{JS}$ and $\epsilon \text{sO}_2$ of the LSTM estimates. D$_\text{JS}$ measures the distance between unpaired samples drawn from two probability distributions $P$ and $Q$. D$_\text{JS}$ is a symmetric version of the Kullback-Leiber divergence~\cite{kullback1951information} (D$_\text{KL}$) and is defined as

\begin{equation}
    \text{D}_\text{JS}(P || Q) = \dfrac{1}{2}\text{D}_\text{KL}(P || M) + \dfrac{1}{2}\text{D}_\text{KL}(Q || M),
\end{equation}

where $M = \dfrac{1}{2}(P + Q)$ and the discrete D$_\text{KL}$ is defined as the relative entropy between two probability distributions:

\begin{equation}
    \text{D}_\text{KL}(P || Q) = \sum\limits_{x\in X} P(x) \log \left( \dfrac{P(x)}{Q(x)}\right).
\end{equation}

To apply these measures, it is important to consider (1) handling the multidimensional probability distributions arising from multi-wavelength measurements and (2) the transformation of two sample distributions into the same sample space. We calculate an aggregate $\overline{D_\text{JS}}$ by calculating the mean over the distance for each wavelength in the spectrum:

\begin{equation}
    \overline{\text{D}_\text{JS}} = \dfrac{1}{N_\lambda}\sum\limits_{\lambda\in\Lambda} \text{D}_\text{JS}(P_\lambda || Q_\lambda),
\end{equation}

where $N_\lambda$ is the number of all available wavelengths $\Lambda$. To standardise the sample space, a z-score normalisation is performed for each spectrum and a histogram with 100 bins ranging from -3$\sigma$ to 3$\sigma$ is created. A Python implementation of the Jensen-Shannon distance, available in the Scipy (v1.10.1) package~\cite{2020SciPy-NMeth}, was used.

\section{Results}
\label{sec:results} 

\subsection{The LSTM-based method achieves accurate results across a range of available input wavelengths.}

\begin{figure}[h!tb]
    \centering
    \includegraphics[width=\textwidth]{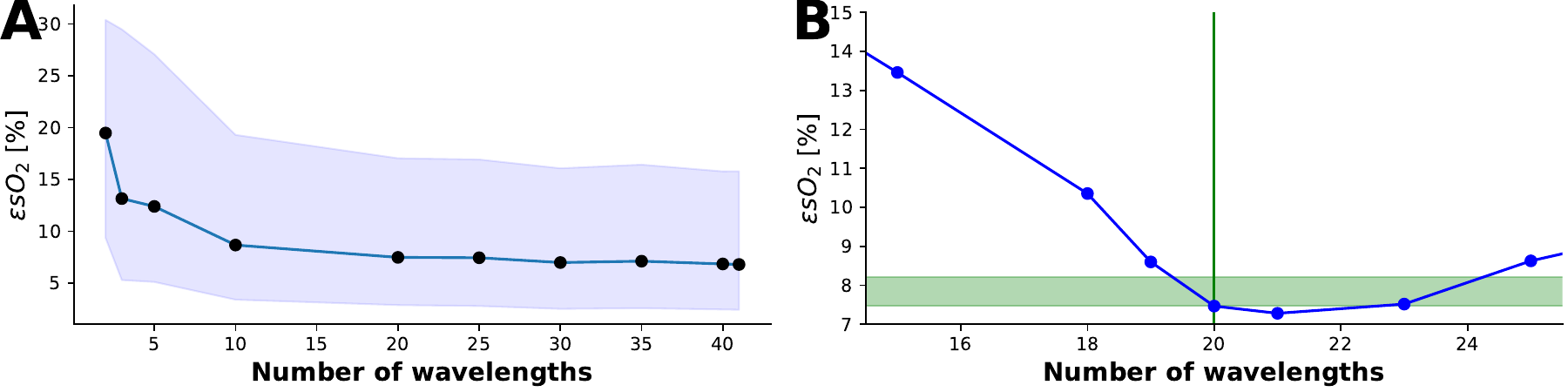}
    \caption{\textbf{The LSTM-based method shows wavelength flexibility.} \textbf{A} LSTMs were trained with varying numbers of wavelengths ($N_\lambda$) to show that with an increasing number of wavelengths, the accuracy of the predictions increases. \textbf{B} LSTM trained at a given $N_\lambda$ can be applied to data with different $N_\lambda$ but yields best results when $N_\lambda$ of the test spectra matches that of the training spectra (indicated by the green vertical line).}
    \label{fig:model_flexibility}
\end{figure}

The accuracy of the LSTM-based network architecture was first tested on the BASE data set, when varying the numbers of available wavelengths ($N_\lambda$) for training or inference. $\epsilon \text{sO}_2$ was extracted when training and testing the network on a certain fixed $N_\lambda$, ranging from 3 to 41 wavelengths (Fig~\ref{fig:model_flexibility} A). We found that $\epsilon \text{sO}_2$ increased as more wavelengths were used for training. Nevertheless, $\epsilon \text{sO}_2$ for $N_\lambda=10$ was only slightly higher than for $N_\lambda=41$ wavelengths. For $N_\lambda<10$, $\epsilon \text{sO}_2$ starts to rapidly increase, which aligns with prior literature~\cite{tzoumas2016eigenspectra} and is potentially exacerbated due to the 'vanishing gradient problem'~\cite{schmidt2019recurrent} in LSTMs, which arises when a substantial portion of the input parameter space consists of zeroes. 

Next, a network trained at a fixed $N_\lambda$ (in this case $N_\lambda=20$) was tested on data with a different $N_\lambda$ (Fig~\ref{fig:model_flexibility} B). Accuracy was found to decrease rapidly if fewer wavelengths were used for testing, but the error remains low if slightly more wavelengths are used. Nevertheless, the results show that the LSTM-based network performs best if the $N_\lambda$ used during inference matches the $N_\lambda$ during training. 
%It should also be noted that we did not encounter performance differences when employing different sampling strategies during training other than random sampling.

\subsection{\textit{In silico} cross-validation reveals the effect of changing simulation parameters.}

\begin{figure}[h!tb]
    \centering
    \includegraphics[width=\textwidth]{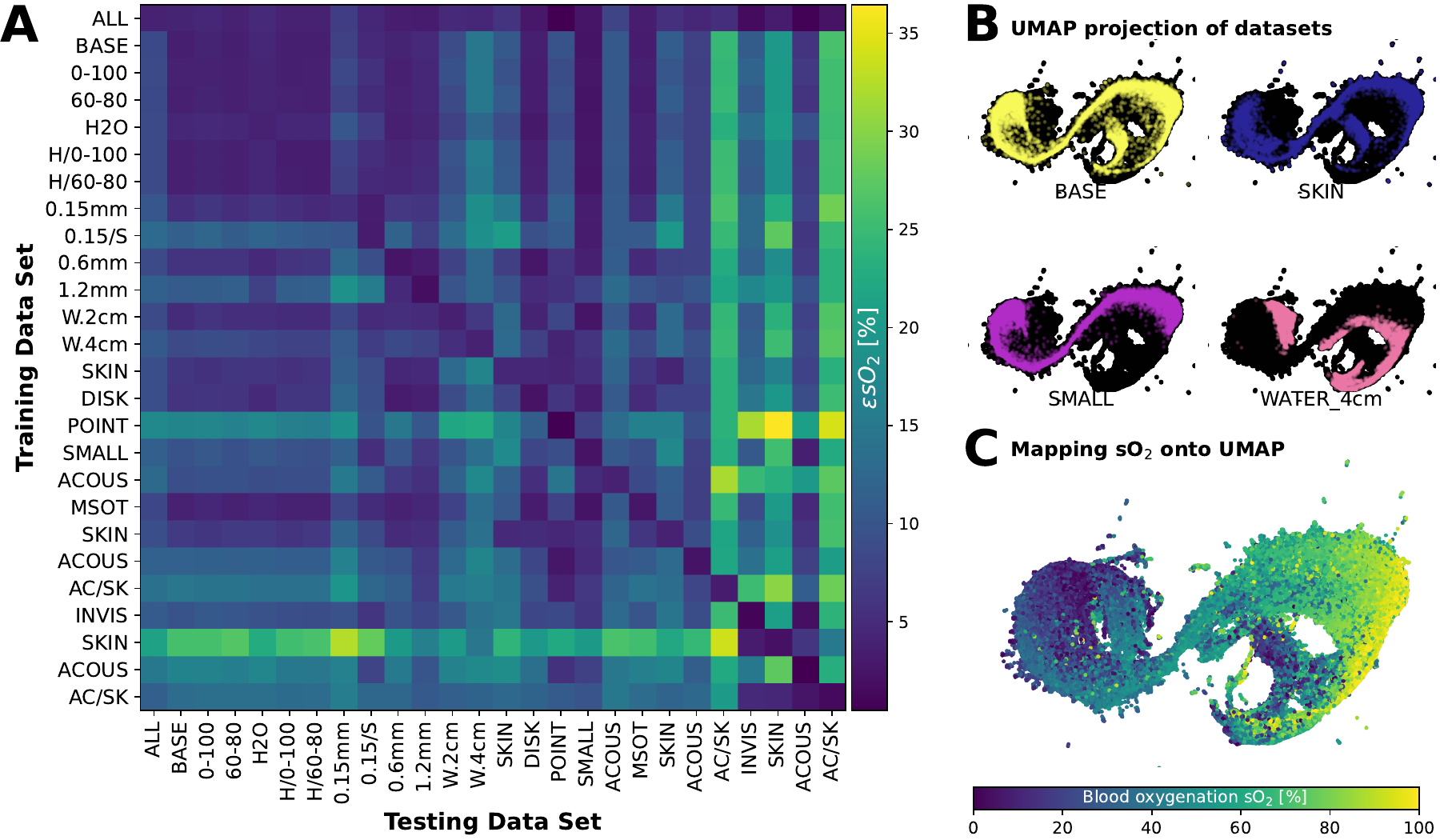}
    \caption{\textbf{Dimensionality reduction and cross-validation reveal systematic differences between training datasets.} \textbf{A} An LSTM-based network trained on each dataset is then applied to every other dataset and all $\epsilon \text{sO}_2$ (median absolute error in percentage points) can be visualised as a performance matrix. Dataset names are shortened for visibility but are detailed in Table \ref{tab:datasets}. \textbf{B} Uniform Manifold Approximation and Projection (UMAP) projections of the four representative example datasets onto an embedding of all training data. \textbf{C} Mapping the ground truth sO$_2$ onto the same projection reveals a correlation along the first UMAP axis.}
    \label{fig:in_silico}
\end{figure}

For each of the 24 simulation parameter variations of BASE, 500 data points with random spatial vessel distributions and a fixed random number generator seed for reproducibility were generated. To investigate the sensitivity of data-driven $\text{sO}_2$ estimates to changes in training dataset parameters, we first trained LSTM-based networks using all 41 available wavelengths on each simulated training dataset and one on a mixed dataset (ALL). We then performed cross-validation on all datasets by applying every trained network to each of the datasets and calculating the median estimation error $\epsilon \text{sO}_2$ (Fig~\ref{fig:in_silico} A). The $\epsilon \text{sO}_2$ values range from 0.5\% to over 35\%. As expected, the best performance occurs when testing on the training set, however, it is important to note that $\epsilon \text{sO}_2$ is not zero (instead ranging from 0.5\% - 5\%), suggesting that the network has not overfitted the training data.

We calculated a Uniform Manifold Approximation and Projection (UMAP)~\cite{mcinnes2018umap} embedding of 200,000 randomly chosen spectra from all datasets. With UMAP, we visualised the location of the spectra of representative datasets on this embedding (Fig~\ref{fig:in_silico} B). Examining this visualisation indicates that some changes in the dataset parameters result in highly different spectra (e.g.: adding a layer of water on top of the tissue), while others lead to only minor variations (e.g.: changing the background oxygenation). Labelling the UMAP embedding with the corresponding ground truth sO$_2$ values reveals a correlation from low to high oxygenation along the horizontal UMAP axis (Fig~\ref{fig:in_silico} C).\\

From the \textit{in silico} cross-validation heatmap (Fig.~\ref{fig:in_silico} A), we can derive several key observations concerning the design of simulated data:

\begin{enumerate}
    \item \textbf{Variation in background sO$_2$ has minimal effect} with the used 1\% blood volume fraction used, however, this could become more significant at higher blood volume fractions.
    \item \textbf{Resolution matters:} Performance improves with higher spatial resolution simulations in the training data, suggesting fine details in the spectral data are important for accurate sO$_2$ estimation.
    \item \textbf{Illumination matters:} When changing from a Gaussian to a point source illumination the error increased.
    \item \textbf{Chromophore inclusion:} When the test dataset contained melanin, but not the training data, the estimation error increased by an average of 5.8 percentage points. When designing a training dataset, all chromophores relevant to the target application should thus be included.
    \item \textbf{Acoustic modelling causes systematic changes:} Using acoustic modelling and image reconstruction introduced systematic spectral changes that increase $\epsilon \text{sO}_2$. This can have detrimental effects on the estimation accuracy and should be considered during training data simulation.
    \item \textbf{Training on a combined dataset is better:} Including random samples from all training datasets yielded more accurate estimates for all test datasets \textit{in silico}. It should already be noted that this finding was not reproducible on the experimental datasets, suggesting that the LSTM-based method was not able to generalise better by training on a combined dataset.
\end{enumerate}

\subsection{The Jensen-Shannon divergence correlates with the estimation error and can therefore be used to identify the best training data set.}

\begin{figure}[h!tb]
    \centering
    \includegraphics[width=\textwidth]{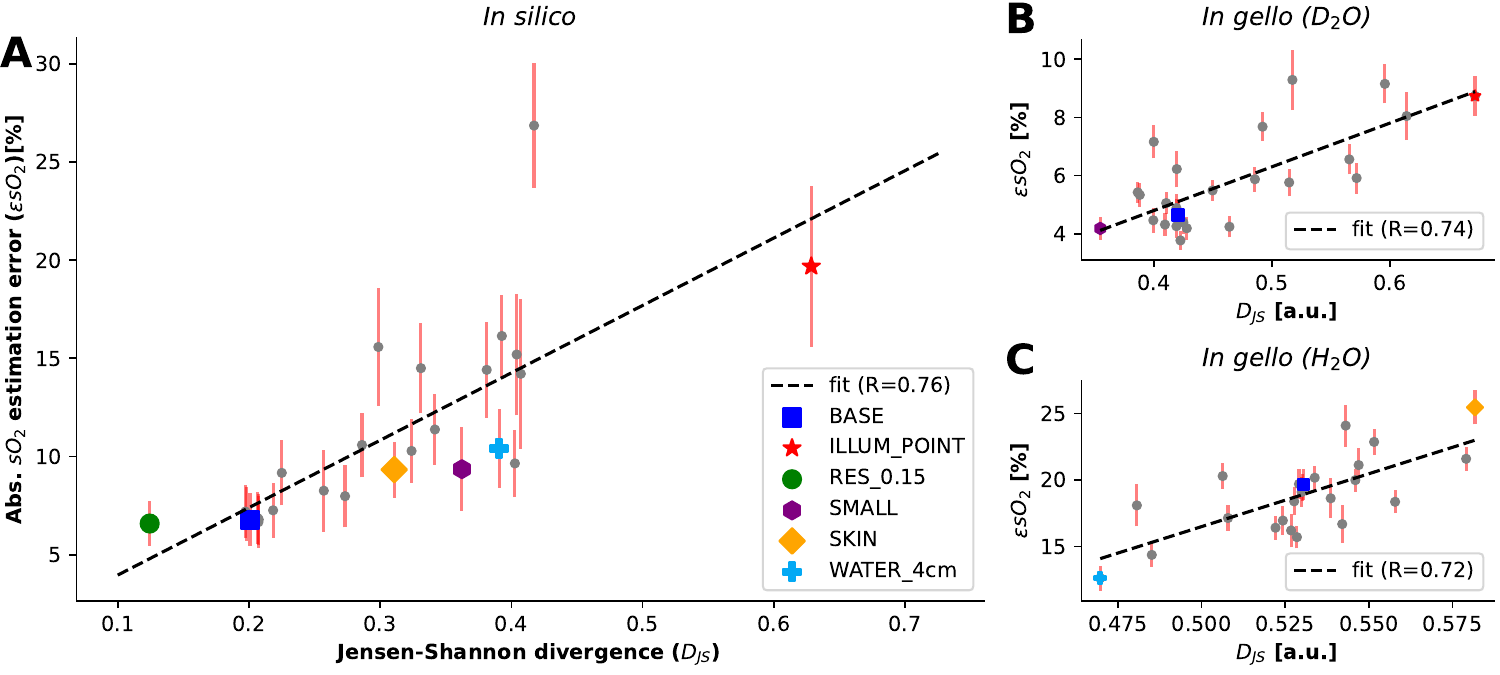}
    \caption{\textbf{The Jensen-Shannon divergence (D$_\text{JS}$) can predict estimation performance.} \textbf{A} D$_\text{JS}$ correlates with the median absolute sO$_2$ estimation error $\epsilon \text{sO}_2$ when applying all networks, each trained on a distinct training dataset, to the baseline dataset (BASE). \textbf{B} D$_\text{JS}$ for the D$_2$O flow phantom data, which shows a similar correlation with $\epsilon \text{sO}_2$. \textbf{C} After removing two outliers, D$_\text{JS}$ shows the same degree of correlation with $\epsilon \text{sO}_2$ for the H$_2$O flow phantom data.}
    \label{fig:jsd_measure}
\end{figure}

Given the variance in performance introduced by the choice of training data, it is desirable to automatically determine the best training dataset for a given algorithm and target application. The Jensen-Shannon divergence (D$_\text{JS}$) allows one to quantify the distance between the data distribution of each dataset and the target data. We calculated the correlation between D$_\text{JS}$ and the median absolute sO$_2$ estimation error ($\epsilon \text{sO}_2$). When applying all networks, each trained on a distinct training dataset, to the BASE dataset, we found that D$_\text{JS}$ correlates strongly with $\epsilon \text{sO}_2$ (Pearson correlation coefficient R=0.76). The RES\_0.15 dataset, which is the dataset simulated at the highest resolution, and not the BASE dataset, achieved the best D$_\text{JS}$ score. This may be because BASE dataset is a subset of the RES\_0.15 dataset, resulting in low $\epsilon \text{sO}_2$ (Fig~\ref{fig:jsd_measure} A).\\

Extending the analysis to experimentally acquired \textit{in gello} D$_2$O flow phantom data showed a similar correlation (R=0.74). The network trained on SMALL achieved the best score with D$_\text{JS} = 0.35$ and $\epsilon sO_2 = 4.2\%$ (Fig \ref{fig:jsd_measure} B). The application of D$_\text{JS}$ to the H$_2$O flow phantom experiment initially revealed no correlation (R=-0.1), however, networks trained on ILLUM\_POINT and MSOT\_ACOUS\_SKIN were outliers and after removing these, the correlation was comparable to other datasets (R=0.72, Fig \ref{fig:jsd_measure} C). The network trained on WATER\_4cm achieved the best score with D$_\text{JS} = 0.47$ and $\epsilon sO_2 = 12.6\%$ (Fig \ref{fig:jsd_measure} B). The presence of outliers emphasises the importance of expert oversight when applying summary metrics such as D$_\text{JS}$.\\

D$_\text{JS}$ correlates with $\epsilon sO_2$ across multiple simulated and experimental data sets, providing evidence that the Jensen-Shannon divergence can predict algorithm performance. This is particularly relevant for previously unseen datasets where the true sO$_2$ is unknown. For each training dataset, a D$_\text{JS}$ value can be computed by drawing random samples from both the training and unseen dataset as outlined in section~\ref{sec:djs}. An LSTM-based network pre-trained on the dataset with the lowest corresponding D$_\text{JS}$ would then be chosen for data analysis since lower D$_\text{JS}$ correlates with a lower $\epsilon sO_2$. The same strategy could also be used to guide an optimisation process to tailor the simulation parameters to create a new training dataset that matches the target application.

\subsection{\textit{In gello} testing shows that the LSTM method outperforms learned spectral decolouring.}

\begin{figure}[h!tb]
    \centering
    \includegraphics[width=\textwidth]{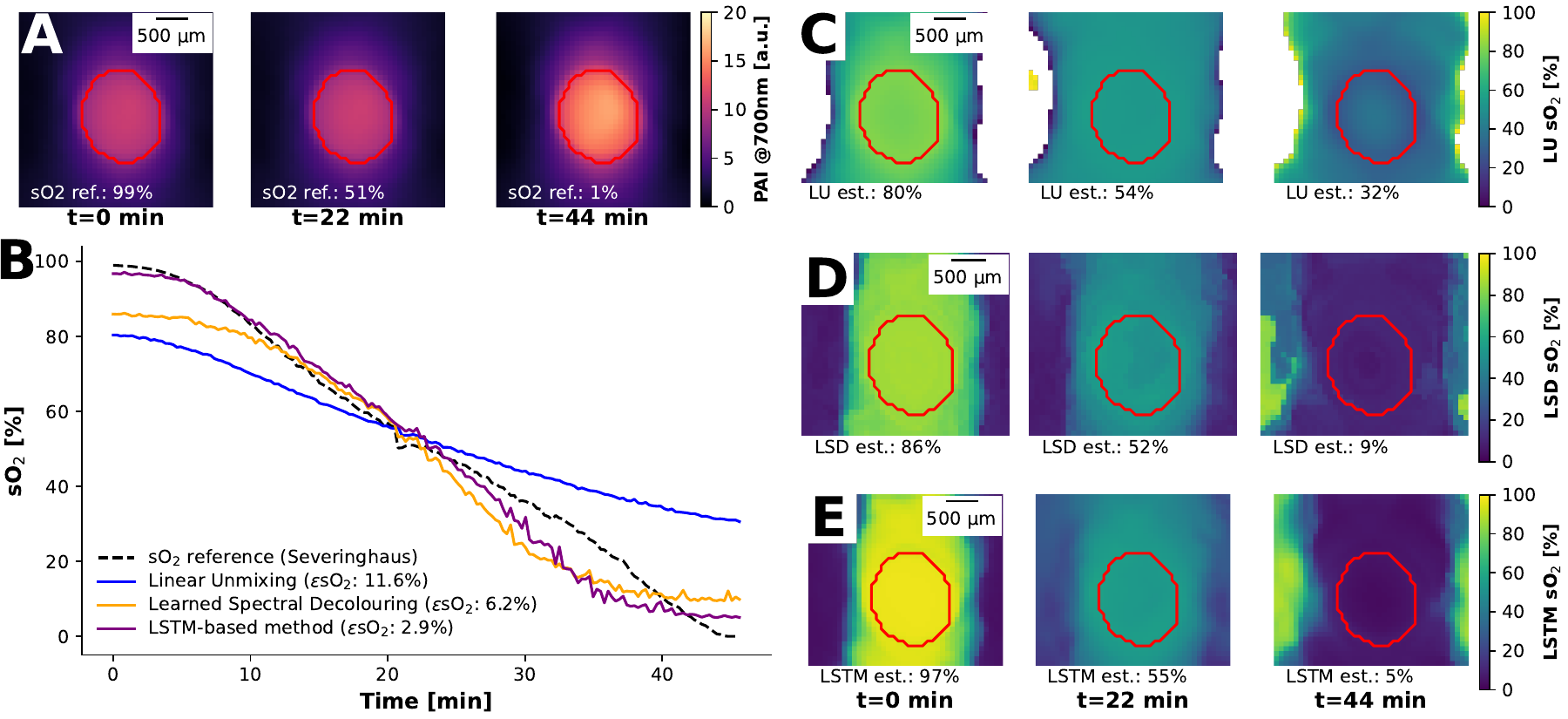}
    \caption{\textbf{Estimation of flow phantom data highlights performance dependence on training dataset.} Three example images of the D$_2$O flow phantom are shown at different time points (0, 22, 44\,min) displaying \textbf{A} the photoacoustic signal intensity at 700\,nm with a red contour marking the blood-carrying tube and \textbf{B} the sO$_2$ estimations from different methods. We visually compare the performance of linear unmixing (LU, \textbf{C}), learned spectral decoloring (LSD, \textbf{D}) and the lstm-based method (LSTM, \textbf{E}) by plotting the sO$_2$ estimations over the same image section and time points shown in \textbf{A}. }
    \label{fig:ingello}
\end{figure}

Algorithm performance on the oxygenation flow phantom was compared with LU as the \textit{de facto} state of the art and with a previously proposed learned spectral decolouring method~\cite{grohl2021learned}. We show three example PA intensity images at 700\,nm of the D$_2$O flow phantom at three time points t=[0\,mins, 44\,mins, 70\,mins] (Fig~\ref{fig:ingello} A), annotated with reference oxygenation (sO$_2^{\text{ref}}$) calculated from $pO_2$ reference measurements using the Severinghaus equation~\cite{severinghaus1979simple}. 

Comparing the estimates of all methods trained on the SMALL training dataset by plotting the estimated sO$_2$ over time (Fig~\ref{fig:ingello} B) reveals that the LSTM-based method is, on average, more than twice as accurate as the learned spectral decolouring (LSD) method and four times as accurate compared to LU. We show the LU estimates for the three example images (Fig~\ref{fig:ingello} C), demonstrating the restricted dynamic range of LU estimates from t=0\,min to t=44\,min, ranging from 80\% to 32\%, compared to a ground truth of 99\% to 1\%. Example images for LSD (Fig~\ref{fig:ingello} D) and the LSTM-based method (Fig~\ref{fig:ingello} E) are shown as well, where the latter can recover the widest dynamic range, extending from 97\% to 5\%. For both methods, we chose SMALL as the training data set, as it was assigned the lowest D$_\text{JS}$ score.

\newpage
\subsection{Static \textit{in vivo} testing shows the applicability of the proposed method to different use cases.}

\begin{figure}[h!tb] 
    \centering
    \includegraphics[width=\textwidth]{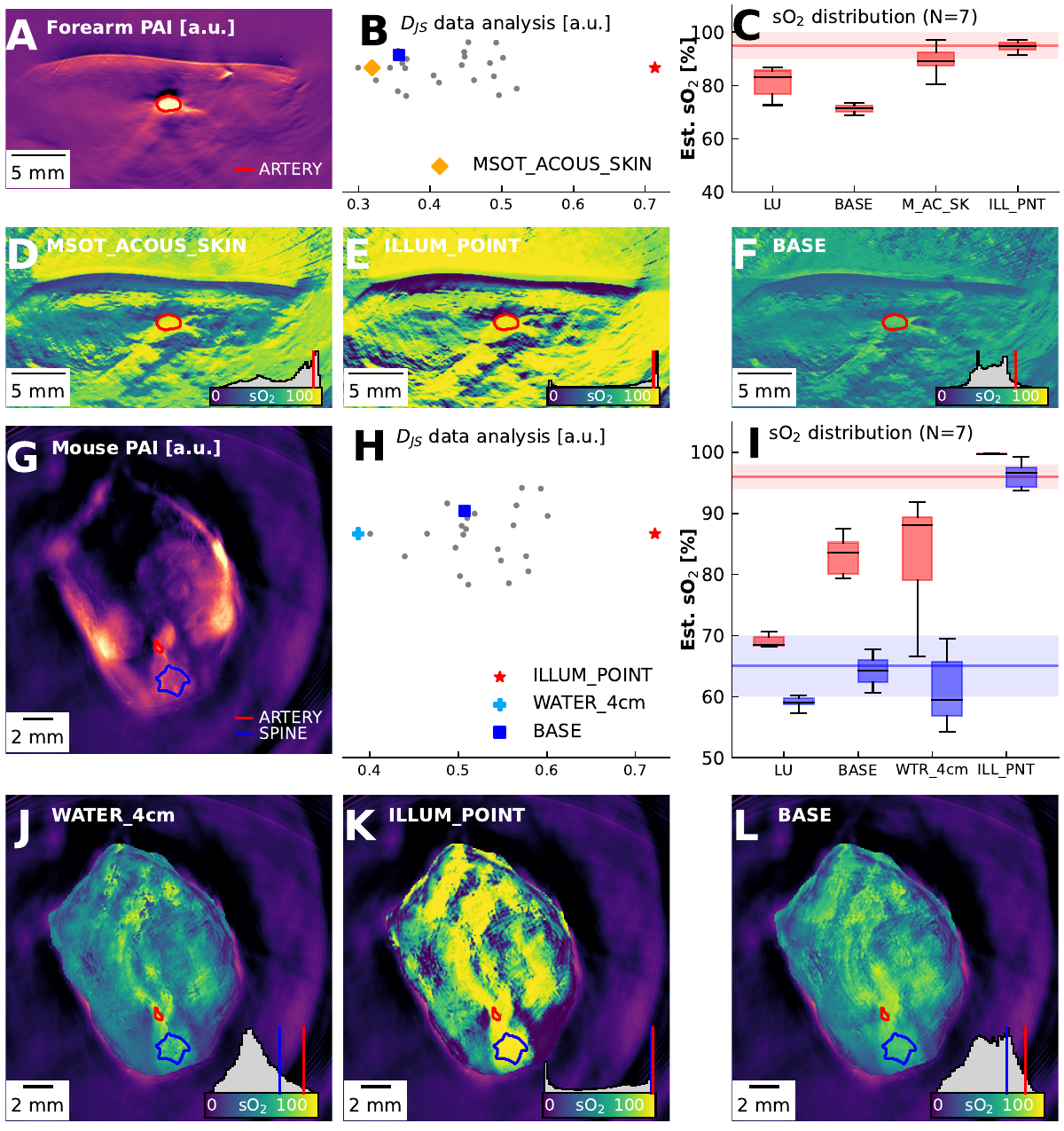}
    \caption{\textbf{Jensen-Shannon divergence (D$_\text{JS}$) proves valuable for \textit{in vivo} data.} LU and LSTM applied to measurements of the human forearm (\textbf{A-F}) and mice abdomens (\textbf{G-L}). Panels \textbf{A} and \textbf{G} show the photoacoustic signal at 800\,nm and panels \textbf{B} and \textbf{H} show the spread of D$_\text{JS}$ estimates for the training datasets. Panels \textbf{C} and \textbf{I} show boxplots of the highlighted regions of interest over all N=7 subjects. The horizontal lines show expected sO$_2$ values for arterial blood (red) and mixed blood (blue). sO$_2$ images are shown for models trained on BASE (\textbf{D, J}) and the best (\textbf{E, K}) and worst (\textbf{F, L}) dataset as predicted by D$_\text{JS}$. On the bottom right of these images, the value distribution is shown as a grey histogram with the mean values of the regions of interest highlighted in their respective colour.}
    \label{fig:invivo}
\end{figure}

Having examined performance in the phantom system with known ground truth, we next apply the data-driven methods to static photoacoustic measurements of seven human forearms (Fig~\ref{fig:invivo} A-F) and seven mouse abdomens (Fig~\ref{fig:invivo} G-L). For each, we: show an example PA image, calculate D$_\text{JS}$ on the data distribution, and compare the estimated blood oxygenation in a region of interest (human forearm: radial artery; mouse: aorta and spine) with literature references.

For the forearm data (Fig~\ref{fig:invivo} A), the MSOT\_ACOUS\_SKIN dataset is objectively the best fit and was assigned the second highest D$_\text{JS}$ score, whereas ILLUM\_POINT dataset was the worst fit (Fig~\ref{fig:invivo} B). Some estimated sO$_2$ values were close to the expected radial artery sO$_2$ value of 95\%-100\%. Notably, the network trained on the MSOT\_ACOUS\_SKIN dataset results in $sO_2\approx$ 90\%, while the network trained on ILLUM\_POINT dataset produces $sO_2\approx$ 95\% (Fig~\ref{fig:invivo} C). 

The sO$_2$ estimates of the network trained on the MSOT\_ACOUS\_SKIN dataset (Fig~\ref{fig:invivo} D) seem to have three primary modes: high values $>$ 80\% from the vessel structures, values in the 60\% to 80\% range in the surrounding tissue, and low values from 10\% to 50\% in the skin and deep tissue. The sO$_2$ estimates of the network trained on the ILLUM\_POINT dataset (Fig~\ref{fig:invivo} E), on the other hand, are concentrated on high sO$_2$ values in all superficial tissue and only seem to be below 85\% in the skin and in deep tissue. The network trained on the BASE dataset (Fig~\ref{fig:invivo} F) estimates low sO$_2$ values throughout the entire tissue and does not exceed 80\%. The ILLUM\_POINT dataset, while seemingly successful if only considering values from the radial artery, was assigned the lowest D$_\text{JS}$ value. The estimates and marginal histograms show that many estimates are mapped to $<$20\% and $>$90\%, explaining the good score in the radial artery. This finding demonstrates a common limitation of data-driven oximetry methods, where the estimated value distributions do not agree with expectations based on human physiology. The combination of all datasets (ALL) results in an extremely low sO$_2$ estimate in the radial artery (median sO$_2$ $<50\%$), which contradicts the \textit{in silico} cross-validation results and indicates overfitting of the method to the training datasets.\\

For mouse images, the aorta and the area around the spinal cord are examined (Fig~\ref{fig:invivo} G), assuming from literature a physiological arterial sO$_2$ of 92\% - 98\% and for the spinal cord, a mixed arterial and venous blood with sO$_2$ of 60\% - 70\%. WATER\_4cm is objectively the best matching dataset and was assigned the highest score according to D$_\text{JS}$ (Fig~\ref{fig:invivo} H). All data-driven methods significantly increase the sO$_2$ estimate in the aorta, and lie within the desired bounds in the spinal cord (Fig~\ref{fig:invivo} I). The BASE (Fig~\ref{fig:invivo} L) and WATER\_4cm (Fig~\ref{fig:invivo} J) datasets estimate a broad distribution and yield a higher sO$_2$ estimate in the aorta and a larger spread between sO$_2$ in the aorta and spinal cord. The limitations of the ILLUM\_POINT (Fig~\ref{fig:invivo} K) dataset are even more evident in the mouse data, where even more pixels are either assigned 0\% or 100\%.

\subsection{Dynamic \textit{in vivo} testing demonstrates that LSTM method can reveal physiological processes.}

To provide a quantifiable decrease in the \textit{in vivo} sO$_2$ levels in mice that could test the capability of the LSTM-based method, we imaged N=6 mice when experiencing asphyxiation breathing 100\% CO$_2$ ~\cite{fadhel2020fluence}. sO$_2$ estimates were extracted from the major visible organs in the scan (spleen, kidneys, spinal cord, and aorta) at 3 minutes before and 10 minutes after CO$_2$ asphyxiation. 

\begin{figure}[h!tb]
    \centering
    \includegraphics[width=\textwidth]{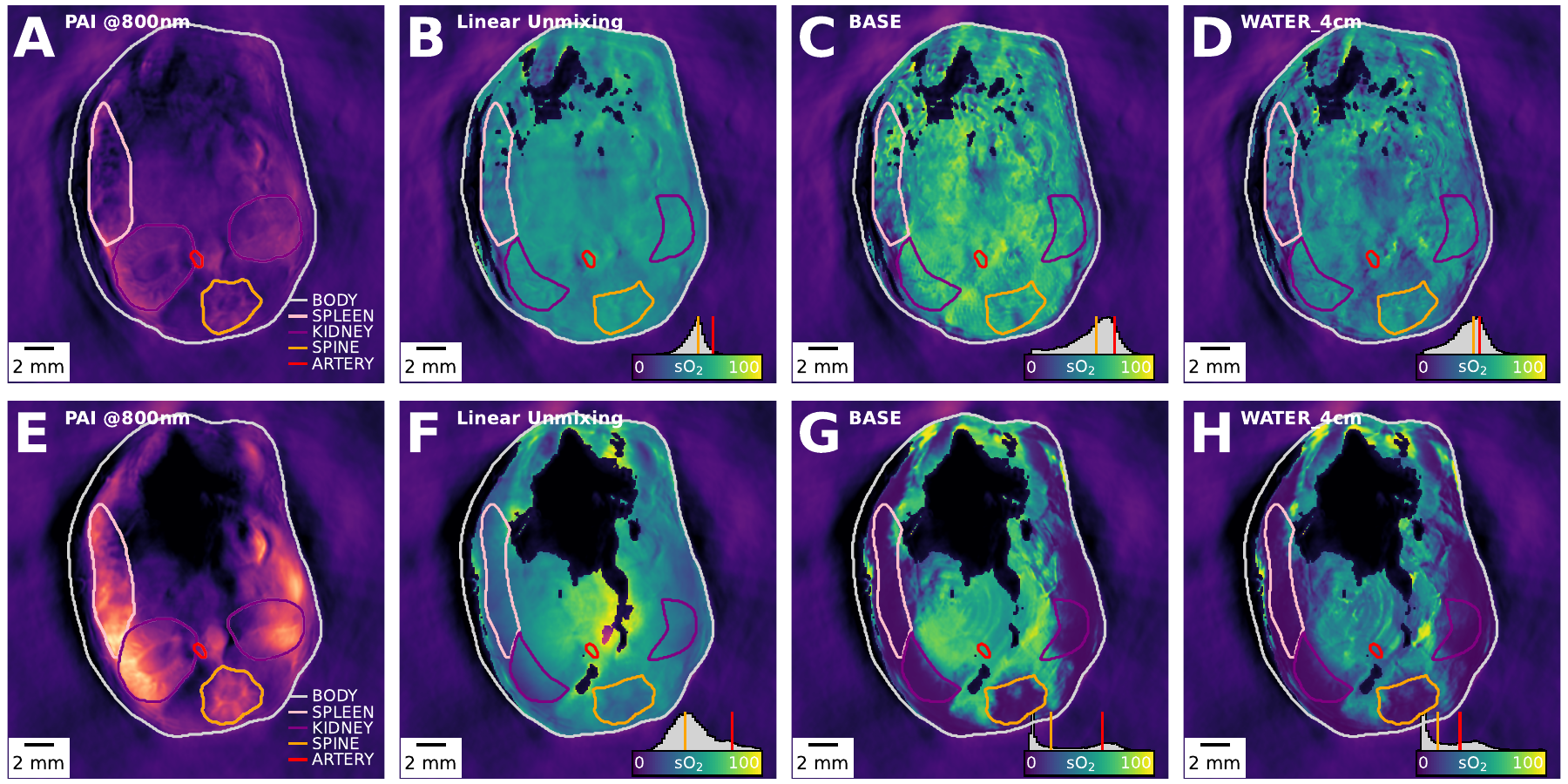}
    \caption{\textbf{Data-driven methods estimate an increased sO$_2$ dynamic range during CO$_2$ delivery compared to linear unmixing.} A single representative mouse is shown here. Panels \textbf{A-D} show the photoacoustic image (\textbf{A}) and sO$_2$ estimation results (\textbf{B-D}) 3 minutes \textit{before} asphyxiation and panels \textbf{E-H} show the photoacoustic image (\textbf{E}) and sO$_2$ estimation results (\textbf{F-H}) 10 minutes \textit{after} asphyxiation. We show sO$_2$ estimates for linear unmixing (\textbf{B, F}), the baseline dataset (BASE, \textbf{C, G}), and the best dataset as predicted by the Jensen-Shannon divergence (WATER\_4cm \textbf{D, H}). Panels \textbf{A} and \textbf{E} show the outlines of the full-organ segmentations, and all other panels \textbf{B-D, F-H} show outlines of the segmented regions used in Table \ref{tab:co2_results}.}
    \label{fig:cO2}
\end{figure}

CO$_2$ asphyxiation (before, Fig~\ref{fig:cO2} A-D; after, Fig~\ref{fig:cO2} E-F) increases the PA signal amplitude at 800\,nm (Fig~\ref{fig:cO2} A,E) in the superficial organs up to a depth of approximately 3\,mm, while the centre of the mouse shows a decrease in signal. sO$_2$ values in the examined organs before CO$_2$ asphyxiation are generally consistent between LU and data-driven unmixing methods, with the network trained on WATER\_4cm estimating slightly lower sO$_2$ values (5 to 8 percentage points lower) (Table ~\ref{tab:co2_results}). Notably, the direction of predicted effects on sO$_2$ levels aligns well between LU and data-driven unmixing methods. Still, the effect sizes are up to three times greater when utilizing data-driven approaches, demonstrating a wider dynamic range. Intriguingly, in the case of the aorta, LU predicts an increase in sO$_2$ levels despite the expected global decrease in sO$_2$ levels due to CO$_2$ exposure. This may be caused by the aforementioned increase in absorption coefficient in the periphery leading to an increase in spectral colouring in depth.

\begin{table}[h!tb]\centering\begin{tabular}{rcccccc}
\textbf{Dataset} $\rightarrow$& \multicolumn{2}{c}{LU} & \multicolumn{2}{c}{BASE} & \multicolumn{2}{c}{WATER\_4cm}\\
 \textbf{Organ} $\downarrow$\hspace*{.5em}& Before [\%] & $\Delta sO_2$ [\%] & Before [\%] & $\Delta sO_2$ [\%] & Before [\%] & $\Delta sO_2$ [\%] \\
 \hline
\\
 Body & 46$\pm$14 & -2 (n.s.) & 44$\pm$20 & -7 (**) & 37$\pm$17 & -8 (**) \\
 Spleen & 45$\pm$10 & -10 (**) & 40$\pm$19 & -24 (**) & 35$\pm$15 & -26 (**) \\
 Kidney & 52$\pm$6 & -14 (**) & 54$\pm$17 & -29 (**) & 47$\pm$11 & -32 (**) \\
 Spine & 55$\pm$6 & -14 (**) & 60$\pm$11 & -35 (**) & 52$\pm$11 & -34 (**) \\
 Aorta & 67$\pm$6 & +5 (n.s.) & 76$\pm$5 & -7 (*) & 68$\pm$17 & -18 (*) \\
 \\
\end{tabular}\caption{\textbf{sO$_2$ decreases during CO$_2$ asphyxiation.} Reported are the mean $\pm$ standard deviation of sO$_2$ measurement \textit{before} CO$_2$ asphyxiation (Before) and the change in the mean ($\Delta sO_2$) \textit{after} the procedure. Values are reported at a maximum depth of 3\,mm into the mouse body for: linear unmixing (LU), an LSTM-based network trained on the baseline training dataset (BASE), and on the best fitting dataset according to the Jensen-Shannon divergence (WATER\_4cm). p-values for $\Delta sO_2$ are calculated using a non-parametric Mann-Whitney U test and indicated by (n.s. = not significant; * = p$<0.05$; ** = p$<0.01$). The segmentation masks are adjusted in Fig~\ref{fig:cO2} \textbf{B-D, F-H} to show the region of interest considered when calculating the results.}\label{tab:co2_results}\end{table}

\section{Discussion}
\label{sec:discussion}

We present an LSTM-based method for estimating sO$_2$ from multispectral PA images. We demonstrate that it can yield superior inference results compared to the previously proposed learned spectral decolouring method while at the same time being usable in a flexible manner, making it a promising candidate to replace LU as the \textit{de facto} state of the art. We also show that the performance of the trained networks is highly dependent on the training data and that changes in simulation parameters can lead to drastically different data distributions. We thus propose to use the Jensen-Shannon divergence (D$_\text{JS}$) to complement the LSTM-based method. D$_\text{JS}$ correlates with the median absolute sO$_2$ estimation error and can thus be used to select the best-fitting training dataset or to optimise the training data distribution to fit the target application.

We highlight how the interplay of the LSTM-based method with D$_\text{JS}$ can be used on a diversity of \textit{in vivo} human and mouse data acquired with different scanners and demonstrate that the LSTM-based method can reveal significant dependencies in sO$_2$ changes that conventional LU would fail to identify. The LSTM-based method further consistently outperforms LU, with estimated sO$_2$ values aligning better with ground truth measurements \textit{in gello} and literature references \textit{in vivo}. LU also shows significant outliers in regions where imaging artefacts are present, which are either not present or less pronounced when using the LSTM-based method.\\

Our \textit{in silico} cross-validation reveals that acoustic modelling and image reconstruction introduces systematic spectral changes not explained by the initial pressure spectrum alone. Thus, an accurate digital model of the clinically used device is crucial during data simulation to ensure the best algorithm performance. While the combined dataset showed promising results \textit{in silico}, these were not replicated in the experimental datasets, which may suggest that the network is overfitting and able to differentiate between the different simulated datasets.\\

D$_\text{JS}$ appears to be a valuable measure for determining the optimal training dataset for the LSTM-based method, as it correlates with the median absolute sO$_2$ estimation error $\epsilon \text{sO}_2$ \textit{in silico} (R=0.76) and \textit{in gello} (R=0.74/0.72). D$_\text{JS}$ predicts a plausible training dataset for all three \textit{in vivo} applications tested in this study, where the predicted value range was 0.4 - 0.7 on the mouse data, 0.5 - 0.8 on the forearm data, and 0.3 - 0.7 on the CO$_2$ data. With the development of fast and auto-differentiable simulation pipelines~\cite{rix2023efficient}, it should be possible to optimise the simulation parameters for accurate sO$_2$ estimates by iteratively minimising D$_\text{JS}$. When using differentiable implementations of distribution distance measures, it might even be possible to integrate this optimisation into an unsupervised training routine.\\

The \textit{in gello} experiments with H$_2$O as the coupling medium had high $\epsilon \text{sO}_2$ errors for all sO$_2$ estimation methods and D$_\text{JS}$ was consistently high. The predicted best training dataset was WATER\_4cm and the worst ILLUM\_POINT, which is consistent with the \textit{in vivo} mouse experiments also having H$_2$O as the coupling medium. Contrary to D$_\text{JS}$ prediction, ILLUM\_POINT has the lowest $\epsilon \text{sO}_2$, resulting in no correlation (R=-0.11); after removing outliers, the correlation was on par with the other experiments (R=0.72). In both cases, the estimation error was lower than predicted by D$_\text{JS}$; while the distributions were different, the trained networks still managed to estimate accurate sO$_2$ values from the training data. Specifically for the ILLUM\_POINT dataset, judging from the quasi-bimodal distributions of the network's estimates on \textit{in vivo} data, appears to have achieved a good agreement with high sO$_2$ values purely by chance. Such outliers are naturally obscured by summary measures such as D$_\text{JS}$, thus expert oversight is recommended.\\

The \textit{in vivo} experiments with CO$_2$ asphyxiation showed an increase in signal amplitude at 800 nm in the periphery with a decrease in signal in the centre of the mouse. These phenomena suggest an overall increase in the absorption coefficient, which could be caused by a range of factors, including: blood coagulation~\cite{su2012optoacoustic}, erythrocyte aggregation~\cite{saha2011simulation}, the presence of bicarbonate ions (HCO$_3^-$) in the blood formed by the dissociation of carbonic acid into bicarbonate and hydrogen ions~\cite{abramczyk2023hemoglobin}, or an increase in blood volume due to blood stasis. Vasoconstriction of the capillaries leading to pallor mortis could also play a role in the better visibility of the superficial organs~\cite{schafer2000colour}. 

Having applied the LSTM-based method combined with D$_\text{JS}$ to a diverse range of data, we believe they provide a promising route to replacing LU for pixel-based PA oximetry. Combining the flexibility in application of LU with the increase in accuracy of deep learning-based unmixing methods is attractive. The CO$_2$ experiment suggests that LU can underestimate the effect size of sO$_2$ changes due to its compressed dynamic range and susceptibility to artefacts. We thus recommend that LU should be complemented by a deep learning-based estimation method. The codes and data of this study are available open-source and the method will be integrated into the PATATO toolkit~\cite{else2023patato}, facilitating its widespread testing and future application.\\ 

Nonetheless, there remain limitations to this study that should be the subject of further research. In this work, we investigated D$_\text{JS}$ predictions on a predefined set of datasets. Based on the results, we believe that D$_\text{JS}$ might be suitable to be used within a minimisation scheme to either manually or automatically determine the best choice of simulation parameters for a given data set, but this remains to be investigated.  To account for spectral colouring artefacts more robustly, the full 2D -- or better 3D -- tissue context should be taken into account within the neural network, as quantitative photoacoustic imaging is only possible with the full 3D context~\cite{cox2012quantitative,grohl2021deep}. Additionally, we have shown that good training data are key for deep learning-based methods for PA oximetry, as such, simulating data as realistically as possible is important. One direction towards this is to make use of domain adaptation methods ~\cite{dreher2023unsupervised,susmelj2023signal,li2022deep} that adapt simulated training data to appear more realistic.

\section{Conclusion}
\label{sec:conclusion} 

The presented LSTM-based approach for sO$_2$ estimation from multispectral PA images surpasses the performance of LU and a previously reported data-driven oximetry method, making it a promising candidate to replace LU as the state of the art. We address the impact of training data variations by introducing the Jensen-Shannon divergence (D$_\text{JS}$) as a valuable complement, enabling selection of optimal datasets and fine-tuning for specific applications. Our LSTM-based method consistently outperforms LU, aligning well with ground truth measurements and literature references, while mitigating outliers in regions prone to imaging artifacts. The combination of the flexibility of the novel LSTM-based method with D$_\text{JS}$ for training data optimisation is a promising direction to make data-driven oximetry methods robustly applicable for clinical use cases.

% \disclosures 
\subsection*{Disclosures}
The authors declare no conflict of interest regarding this work.

\subsection* {Code, Data, and Materials Availability} 

The data and code to reproduce the findings of this study are openly available.\\

The data is available under the CC-BY 4.0 license at: 
\url{https://doi.org/10.17863/CAM.105987}\\

The code is available under the MIT license at:
\url{https://github.com/BohndiekLab/LearnedSpectralUnmixing}\\

\subsection* {Acknowledgments}
This work was funded by: Deutsche Forschungsgemeinschaft [DFG, German Research Foundation] (JMG; GR 5824/1); Cancer Research UK (SB, TRE; C9545/A29580); Cancer Research UK RadNet Cambridge (EVB; C17918/A28870); Against Breast Cancer (LH); the Engineering and Physical Sciences Research Council (SB, EP/R003599/1) %NOTE
The work is supported by the NVIDIA Academic Hardware Grant Program and utilised two Quadro RTX 8000.
The authors would like to thank Dr Mariam-Eleni Oraiopoulou for the helpful discussions.\\

This work was supported by the International Alliance for Cancer Early Detection, a partnership between Cancer Research UK (C14478/A27855), Canary Center at Stanford University, the University of Cambridge, OHSU Knight Cancer Institute, University College London and the University of Manchester.\\

This research was supported by the NIHR Cambridge Biomedical Research Centre (NIHR203312). The views expressed are those of the authors and not necessarily those of the NIHR or the Department of Health and Social Care.\\

For the purpose of open access, the authors have applied a Creative Commons Attribution (CC BY) licence to any Author Accepted Manuscript version arising.

%%%%% References %%%%%
\bibliography{report}   % bibliography data in report.bib
\bibliographystyle{spiejour}   % makes bibtex use spiejour.bst

%%%%% Biographies of authors %%%%%

\textbf{Janek Gr\"ohl} completed his M.Sc. degree in medical computer science in 2016 and completed his PhD in April 2021 under Prof. Lena Maier-Hein at the German Cancer Research Center. Janek is working as a postdoctoral fellow with Prof. Sarah Bohndiek funded by the Walter Benjamin Programme of the German Research Foundation focusing on using machine learning methods and physical modelling to tackle the problem of quantitative photoacoustic imaging.\\

\textbf{Kylie Yeung} completed her MRes in Connected Electronic and Photonic Systems in 2022, jointly between University College London and the University of Cambridge, during which she investigated the effect of different simulated training datasets for quantitative photoacoustic imaging. She is currently pursuing a PhD at the University of Oxford.\\

\textbf{Kevin Gu} completed his MSci degree in experimental and theoretical physics in 2022 at the University of Cambridge. Kevin’s dissertation was focussed on a wavelength-flexible approach to data-driven photoacoustic oximetry. He now works as a quantitative researcher in the Netherlands.\\

\textbf{Thomas R. Else} completed his MSci degree in experimental and theoretical physics in 2019 and is currently pursuing a PhD in medical sciences, both at the University of Cambridge. His research focuses on clinical translation of photoacoustic imaging, with a focus on equitable application of the technology through open access software and evaluation of skin tone biases.\\

\textbf{Monika Golinska} completed her PhD in oncology at the University of Cambridge in 2012. During her postdoc at Prof Sarah Bohndiek’s lab, she studied the relationship between sex hormones and tumour microenvironment in breast cancer. Monika is currently a MCSA fellow at the Medical University of Lodz and researches endometriosis and its link with ovarian cancer.\\

\textbf{Ellie V. Bunce} completed her MPharmacol degree (Hons) at the University of Bath in 2021. She is currently pursuing a PhD in medical sciences at the University of Cambridge, investigating the potential of vascular targeted therapies to enhance cancer radiotherapy response using novel imaging approaches.\\

\textbf{Lina Hacker} is a Junior Research Fellow at the Department of Oncology at the University of Oxford, UK. Her research is focused on the medical and technical validation of novel approaches for cancer imaging, specifically relating to tumour hypoxia. She received her PhD degree in Medical Sciences at the University of Cambridge, UK, and holds a Master’s and Bachelor’s degree in Biomedical Engineering and Molecular Medicine, respectively.\\

\textbf{Sarah E. Bohndiek} completed her PhD in radiation physics at the University College London in 2008 and then worked in both the United Kingdom (at Cambridge) and the United States (at Stanford) as a postdoctoral fellow in molecular imaging. Since 2013, she has been a Group Leader at the University of Cambridge, where she is jointly appointed in the Department of Physics and the Cancer Research UK Cambridge Institute. She was appointed as full professor of biomedical physics in 2020. Sarah was recently awarded the CRUK Future Leaders in Cancer Research Prize and SPIE Early Career Achievement Award in recognition of her innovation in biomedical optics.\\

\end{spacing}
\end{document}